\documentclass[preprint,superscriptaddress,showpacs,nofootinbib]{revtex4}

\usepackage{amssymb}
\usepackage{amsmath}
\usepackage{graphicx}
\usepackage{latexsym}

\newcommand{\GeV}{\text{GeV}}

\begin{document}

\title{Neutralino Dark Matter in Light Higgs Boson Scenario}

\author{Masaki Asano}
\affiliation{Theory Group, KEK, Tsukuba, Ibaraki 305-0801, Japan}
\affiliation{The Graduate University for Advanced Studies (Sokendai), Tsukuba, Ibaraki 305-0801, Japan}
\author{Shigeki Matsumoto}
\affiliation{Inst.\ for Int.\ Adv.\ Interdisciplinary Research, Tohoku University, Sendai, Miyagi 980-8578, Japan}
\author{Masato Senami}
\affiliation{ICRR, University of Tokyo, Kashiwa, Chiba 277-8582, Japan}
\affiliation{Department of Micro Engineering, Kyoto University, Kyoto 606-8501, Japan}
\author{Hiroaki Sugiyama}
\affiliation{SISSA, via Beirut 2-4, I-34100 Trieste, Italy}

\begin{abstract}
 Phenomenology of neutralino dark matter
in the minimal supersymmetric model is discussed
for a scenario where the lightest Higgs boson mass is
lighter than $114.4\,\GeV$.
 We show that the scenario is consistent
not only with many collider experiments
but also with the observed relic abundance of dark matter.
 The allowed region may be probed by experiments
of $B_s \to \mu^+\mu^-$ in near future.
 The scenario predicts a large scattering cross section
between the dark matter and ordinary matter
and thus it may be tested in present direct detection experiments
of dark matter.
\end{abstract}

\pacs{95.35.+d, 12.60.Jv, 14.80.Cp, 13.66.Fg}

\maketitle 
\begin{center}
{\bf I. \ INTRODUCTION}
\end{center}

The Large Hadron Collider (LHC) experiments will start soon.
The primary goal of the LHC experiments is the search for the Higgs boson $\phi$,
which is the last undiscovered particle in the standard model 
of the particle physics (SM). 
The Higgs boson is the only scalar particle in the SM
and rules masses of the other particles via couplings with them.
Hence, the discovery of the Higgs particle is indispensable
to understand the nature.

The present bound on the Higgs mass in the SM comes from the LEP2 experiments
and is given as~\cite{LEPSM}
\begin{eqnarray}
m_\phi > 114.4\,\GeV ~~~ (95\,\%\,\text{C.L.})\, .
\label{eq:SMHiggsbound}
\end{eqnarray}
In addition to the bound,
the Higgs mass should be within TeV scale theoretically
because of the partial wave unitarity of
weak gauge boson scatterings.
 On the other hand, once quantum correction to the Higgs mass square is taken into account,
it has quadratic divergence.
Hence, the natural scale of the SM Higgs mass is the cutoff scale of the SM,
which is, for example, the grand unification theory (GUT) scale
($M_G \sim 10^{16}\,\GeV$) or the Planck scale ($\sim 10^{19}\,\GeV$).
This problem in the SM is called as the hierarchy problem and considered to be very serious.
To solve this problem has been the paradigm for particle phenomenology
for a couple of decades.

The supersymmetry (SUSY)~\cite{BookDrees} is one of solutions to this problem.
The quadratic divergence is canceled
between loop diagrams of SM particles and those of SUSY partners.
Moreover, SUSY theory also has some advantages;
the lightest SUSY particle (LSP) is provided
as a candidate for dark matter with R-parity conservation,
and the grand unification is improved in comparison with the SM\@.
By these reasons, SUSY theory is the most celebrated candidate
for physics beyond the SM\@.

In the minimal SUSY SM (MSSM),
two Higgs doublet fields ($H_u , H_d$) are required
to provide mass terms for quarks and leptons. 
Hence, the MSSM predicts three neutral and two charged Higgs bosons:
two CP-even scalar ones $h$ (lighter), $H$ (heavier),
one CP-odd pseudoscalar one $A$, and a pair of charged ones $H^\pm$.
The mass of the lightest one, $m_h$, can be lighter than
the SM bound~(\ref{eq:SMHiggsbound})
if $g_{ZZh}$, which is the coupling of $h$ to the $Z$ boson, is significantly
smaller than the SM value~\cite{LEPSUSY,LHSBottino1,LHSKane}.
We call the MSSM with the Higgs boson whose mass is lighter than $114.4\,\GeV$
as the light Higgs boson scenario (LHS).
The LHS might explain the $2.3\,\sigma$ level excess of events at
$98\,\GeV$ in the LEP2 experiments~\cite{LEPSM,LEPSUSY}
though it is very difficult to be explained
in the SM with $m_\phi = 98\,\GeV$.

A reference model widely used to study the MSSM
is the constrained MSSM (CMSSM), which is parametrized by $m_0,m_{1/2}$
(universal scalar and gaugino masses at $M_G$),
$A_0$ (universal trilinear coupling at $M_G$),
$\tan \beta \equiv \langle H_u \rangle / \langle H_d \rangle ,$
and $ {\rm sign} (\mu)$.
Unfortunately, the LHS cannot be realized in the CMSSM.
Thus, studies for the LHS has been restricted in spite of
its potential importance~\cite{LHSBottino1,LHSKane,LHSDrees,LHSBelyaev,LHSKim,LHSBottino2}.
In recent works of the LHS,
phenomenological aspect with $m_h \simeq 98\,\GeV$~\cite{LHSDrees}
and $m_h < 90\,\GeV$~\cite{LHSBelyaev} 
and a solution to the little hierarchy problem~\cite{LHSKim} are discussed.

In this Letter, we investigate phenomenology of neutralino dark matter in the LHS.
In particular, it is studied
whether the LSP in the LHS is a viable dark matter candidate,
that is, 
whether the parameter region where the relic abudance of the LSP is consistent
with the observed abundance by the WMAP experiment~\cite{WMAP} is allowed
by the recent severe constraints from $b \to s \gamma$ and $B_s \to \mu^+ \mu^-$.
We calculate the relic abundance 
by using the SUSY model with non-universal scalar masses for the Higgs multiplets (NUHM),
which is the simplest example realizing the LHS and consistent with GUT\@.
We also calculate the cross section of the dark matter with ordinary matter
and compare with present bounds by direct detection experiments of dark matter.

\begin{center}
 {\bf II. \ LIGHT HIGGS BOSON SCENARIO}
\end{center}
The mass eigenstates for the CP-even neutral Higgs bosons in the MSSM are given by
\begin{eqnarray}
\left(
\begin{array}{c}
h  \\ 
H
\end{array}
\right) = 
\left(
\begin{array}{cc}
- \sin \alpha & \cos \alpha   \\ 
\cos \alpha &  \sin \alpha 
\end{array}
\right)
\left(
\begin{array}{c}
{\rm Re} \, H_d^0  \\ 
{\rm Re} \, H_u^0
\end{array}
\right) ,
\end{eqnarray}
where $ \alpha $ is the angle
diagonalizing the mass square matrix of the Higgs bosons,
\begin{eqnarray}
\left(
\begin{array}{cc}
m_A^2 s_\beta^2 + m_Z^2 c_\beta^2 + \Delta_{dd} 
	& - ( m_A^2 + m_Z^2 ) s_\beta c_\beta + \Delta_{du}\\ 
- ( m_A^2 + m_Z^2 ) s_\beta c_\beta + \Delta_{du}
	& m_A^2 c_\beta^2 + m_Z^2 s_\beta^2 + \Delta_{uu}
\end{array}
\right) .
\end{eqnarray}
Here, $m_A$ ($m_Z$) is the pseudoscalar Higgs boson ($Z$ boson) mass,
$c_\beta \equiv \cos \beta, s_\beta \equiv \sin \beta$, and
$\Delta_{ii}$ denotes the radiative correction,
which is given in Ref.\,\cite{BookDrees}.

The LEP collaborations have analyzed their data using several benchmark scenarios
to search Higgs bosons in the MSSM~\cite{LEPSUSY}.
In their results, it is implied that $m_h < 114.4\,\GeV$ is allowed 
if $g_{ZZh}$ is significantly suppressed
since the primary mode to search $h$ in the LEP2 experiments is $e^+ e^- \to Zh$.
The coupling $g_{ZZh}$ is given as
$g_{ZZh} = g_{ZZ\phi} \sin (\beta - \alpha)$,
where $g_{ZZ\phi}$ is the coupling of $\phi$ to $Z$.
To avoid the constraint (\ref{eq:SMHiggsbound}),
so small $\sin (\beta - \alpha)$ is not needed.
When $90\,\GeV < m_h < 114.4\,\GeV$,
the Higgs boson $h$ with $\sin (\beta - \alpha) < 0.5 $ can evade
Higgs boson searches by the LEP2 experiments.
For this suppression, a large mixing $\alpha$ between $h$ and $H$
is required to cancel $\beta$, and hence $m_H \sim m_h$.
Note that $m_H > 114.4\,\GeV$ is also required because $H$ has not been
discovered yet.
On the other hand,
suppressed $ \sin (\beta - \alpha) $ leads unsuppressed $ \cos (\beta - \alpha)$.
This results in large $g_{ZAh}$, since $g_{ZAh} \propto \cos (\beta - \alpha)$.
Therefore, we should be also careful in $e^+ e^- \to Ah$ mode in the LHS\@.
However, since the $g_{ZAh}$ coupling originates in a derivative coupling,
no additional constraint appears for $m_h > 90\,\GeV$
due to P-wave suppression~\cite{LEPSUSY}.

The LHS cannot be realized in the CMSSM as mentioned above.
If $m_0$ and $m_{1/2}$ are fixed to derive $m_h < 114.4\,\GeV$,
masses of other SUSY particles are too small.
For realization of the LHS, the masses of the Higgs doublets
should be different from others as $m_{H_u(H_d)}^2 = (1+\delta_{H_u(H_d)}) m_0^2$.
 It is reasonable
because the Higgs multiplets are not necessarily
in the same multiplet of GUT with other scalar particles.
The simplest model with this boundary condition is the NUHM\@.
Hence, we adopt this model as a reference model to investigate the LHS\@.
The NUHM has six free parameters,
$(m_0, \, m_{1/2}, \, A_0 , \, \tan \beta, \, \mu, \, m_A )$,
where $(m_0, \, m_{1/2}, \, A_0)$ are defined at $M_G$
and the other is defined at the electroweak scale.
This parametrization allows two Higgs doublets to take their masses  
at $M_G$ as free parameters.
Using these values at $m_Z$ scale,
a boundary condition at $M_G$ is derived by the renormalization group running.
Then, masses of SUSY particles run back from $M_G$ to $m_Z$.
In this work, the renormalization group running is evaluated by ISAJET 7.75~\cite{ISAJET}.

\vspace*{10mm}
\begin{center}
 {\bf III. \ CONSTRAINTS}
\end{center}
 We consider the following constraints from collider experiments.

(i) LEP2 $Zh/ZH $ constraints:
In order to satisfy this constraints,
$\sin (\beta - \alpha) < 0.5$ is required
for $90\,\GeV < m_h < 114.4\,\GeV$.
In our analysis,
this constraint is imposed in a more precise form dependent on $m_h$
with assuming that the branching ratios of $h$ decays
are the same as those of the SM~\cite{LEPSUSY}.
 This assumption is good enough
because SUSY loop corrections
to the Yukawa coupling of $b$ quark are small
in our analysis in which we take $\tan\beta = 10$
in order to satisfy the constraint from
$B_s \to \mu^+ \mu^-$ process (See (v) also).
In addition, $m_H> 114.4\,\GeV$ is required 
though $m_H$ cannot be very different from $m_h$.
Even for $m_h < 90\,\GeV$,
the Higgs boson can evade the LEP2 searches
with smaller $\sin (\beta - \alpha) \sim 0.2$.
However, this region is very restricted, and therefore
it is ignored in this work for simplicity.

(ii) LEP2 $Ah/AH $ constraints:
In our calculation, the constraint on $g_{ZAh}$ 
depends on the masses $m_A$ and $m_h$~\cite{LEPSUSY}.
We assumed $m_h$-max benchmark for the final states of the Higgs bosons decay.
Since the dominant decay mode in our case is $b$ quark pair production,
which is the same as the $m_h$-max benchmark,
the uncertainty from this assumption is  small enough.
For $90\,\GeV < m_h < 114.4\,\GeV$,
this constraint is not severe due to the P-wave suppression.

(iii) SUSY particle search constraints~\cite{PDG}:
We consider the constraints for masses of SUSY particles 
through micrOMEGAs 1.3.7 package~\cite{micromegas}
which is also used to calculate the following two processes
and the abundance of the LSP in the LHS\@.
In particular, the constraint for the lightest chargino mass is precisely given
depending on the lightest neutralino and sneutrino masses.

\begin{figure}[t]
\begin{center}
\scalebox{0.9}{\includegraphics*{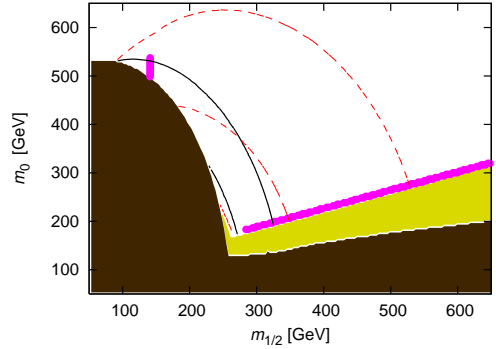}}
\caption{The magenta (gray) region is consistent
with observed dark matter abundance
for $\mu = 600\,\GeV$, $A_0 = -1000\,\GeV$, $\tan \beta = 10$, and $m_A = 100\,\GeV$. 
 See the text for the explanations about other regions and lines.}
\label{fig:omegah2}
\end{center}
\end{figure}

(iv) $b \to s \gamma$ process:
For the small charged Higgs boson mass $m_{H^\pm}$,
the constraint from ${\rm Br} ( b \to s \gamma)$ is serious,
since ${H^\pm}$ contribute constructively
to the SM prediction~\cite{Gambino:2001ew}.
The charged Higgs boson mass is predicted as the electroweak scale in the LHS\@.
This is because generally $m_{H^\pm} \sim m_H \sim m_A$ in the MSSM
and $m_H \sim m_h$ is required for large mixing.
Hence, the contribution from ${H^\pm}$
should somewhat be canceled by SUSY contribution in the LHS~\cite{LHSKim}.
In particular, the $H^\pm$ contribution can be compensated by large
$A$-terms%
~\footnote{
 Large $A$-terms may cause the problem of
charge (and color) breaking minima~\cite{Kusenko:1996jn}.
 We checked that values of $A_0$ used in our analysis
are acceptable because the MSSM minimum
is stable enough even if it is not the global one.
}.
The present experimental bound 
is reported as~\cite{Barberio:2007cr}
\begin{eqnarray}
{\rm Br} ( b \to s \gamma) = (355 \pm 24^{+9}_{-10} \pm3) \times 10^{-6} .
\label{bound_bsg}
\end{eqnarray}
 In our analysis~\footnote{
The recent theoretical prediction of ${\rm Br} ( b \to s \gamma)$ in the SM
is significantly lowered as
${\rm Br} ( b \to s \gamma) = (2.98 \pm 0.26) \times 10^{-4}$~\cite{Becher:2006pu}.
This improvement was not included in our calculation
which gives ${\rm Br} ( b \to s \gamma) = 3.62 \times 10^{-4}$
as the SM prediction.
 Since the deviation of the new value from
the present experimental result (\ref{bound_bsg})
is larger than that of the value we used,
it will shift the allowed region we obtained
for ${\rm Br} ( b \to s \gamma)$
to a heavier mass region in $(m_{1/2}$, $m_0)$ plane
(See also Fig.\,\ref{fig:omegah2})
and/or enable us to use smaller $\tan\beta$.},
we search allowed regions for
\mbox{$304\times 10^{-6} < {\rm Br}(b \to s \gamma ) < 406\times 10^{-6}$}.

(v) $B_s \to \mu^+ \mu^-$ process:
The process, $ B_s \to \mu^+ \mu^- $,
is mediated by neutral Higgs bosons.
The $H/A$ contribution to $ B_s \to \mu^+ \mu^- $ is proportional to $ m_A^{-4}$.
In the LHS, this contribution is large since $ m_A \sim m_H \sim m_h $.
Moreover, this contribution is proportional to $(\tan \beta)^6$.
As a result, the LHS with large $\tan \beta \gtrsim 20$ is almost forbidden.
The present experimental bound is given by~\cite{Mack:2007ra}
\begin{eqnarray}
{\rm Br} ( B_s \to \mu^+ \mu^-) < 5.8 \times 10^{-8} ~~~ ({\rm 95\,\%\,C.L.}) .
\label{eq:Bsmumu}
\end{eqnarray}
 This upper bound is used for our calculation.

\begin{center}
 {\bf IV. \ RESULTS FOR THE DARK MATTER}
\end{center}
 The purpose of this work is not to investigate all parameter space exhaustively
but to indicate that the LHS is
consistent with dark matter abundance observed
by the WMAP experiment~\cite{WMAP},
\begin{eqnarray}
\Omega_{\text{DM}} h^2
 = (0.1277^{+0.0080}_{-0.0079}) - (0.02229 \pm 0.00073 ),
\end{eqnarray}
where $h=0.73^{+0.04}_{-0.03}$ stands for the reduced Hubble constant~\cite{WMAP}.
We adopt the allowed region as
$0.0895 < \Omega_{\rm DM} h^2 < 0.1214$.
 First, we take $ \tan \beta = 10 $.
Small $\tan \beta$ is favored by 
the $ B_s \to \mu^+ \mu^- $ process.
 On the other hand, too small values of $\tan\beta$ are disfavored
by the $b \to s \gamma $ process with moderate values
of $A_0$ for the stable MSSM vacuum.
 Furthermore, we choose $\mu = 300, 600, 700\,\GeV$,
where only positive $ \mu $ is considered because it is favored by muon $g-2$.
The values of $A_0$ are chosen so as to cancel 
the $H^\pm$ contribution to $b \to s \gamma$.
Then, we scan the parameter region where $50\,\GeV < m_0 , m_{1/2} <700\,\GeV$
and $m_A = 80 \text{-} 140\,\GeV$.
This range of $m_A$ is a typical one in the LHS~\cite{LHSKim,LHSBelyaev}.

In Fig.\,\ref{fig:omegah2}, we fix $m_A = 100\,\GeV$ for simplicity.
The region consistent with the dark matter abundance observed
by the WMAP experiment
is shown for $\mu = 600\,\GeV$ and $A_0 = -1000\,\GeV$ in $(m_{1/2}, m_0)$ plane.
The magenta (gray) region is consistent with the observed abundance.
The strip at lower-right portion is the stau coannihilation region
and the vertical strip around $m_{1/2} = 150\,\GeV$
is the $s$-channel pseudoscalar Higgs region.
In the former region,
the abundance of the LSP is decreased sufficiently
by the coannihilation with stau
because of the small difference between their masses.
Note that the difference is about 10\% of their masses
and only a mild tuning is required.
In the latter region,
the LSP annihilation is governed by the pseudoscalar mediated diagram.
It is different from the funnel region
where the LSP annihilation picks up the pseudoscalar Higgs pole,
because $m_A$ is smaller than twice of the LSP mass.
Due to the smallness of $m_A$,
the annihilation cross section of the LSP is sufficiently large even for large $m_0$.
The yellow (light gray) region is not favored
because a charged particle is lighter than the neutralino.
The brown (black) region is 
forbidden by the appearance of tachyonic particles
at the electroweak scale due to large $A_0$.
The prediction for ${\rm Br}(b \to s \gamma )$ is denoted by solid lines,
$(406, 304) \times 10^{-6}$ from top to bottom.
It is seen that the constraint from the $b \to s \gamma $ process
removes broad parameter space.
The cancelation between contributions of SUSY particles and $H^\pm$ 
constrains $m_0$ and $m_{1/2}$ from both above and below.
This is a salient feature in contrast with other SUSY models.
The dashed lines shows
$ {\rm Br} ( B_s \to \mu^+ \mu^-) = (1.5, 2.5, 4) \times 10^{-8}$
from top to bottom.
The recent bound (\ref{eq:Bsmumu}) does not constrain
the magenta (gray) region.
Experiments of $B_s \to \mu^+ \mu^-$ may, however,
probe large part of the region in the future
because the sensitivities are expected to be
better than ${\rm Br} (B_s \to \mu^+ \mu^-) = 1\times 10^{-8}$.

\begin{figure}[t]
\begin{center}
\scalebox{1.15}{\includegraphics*{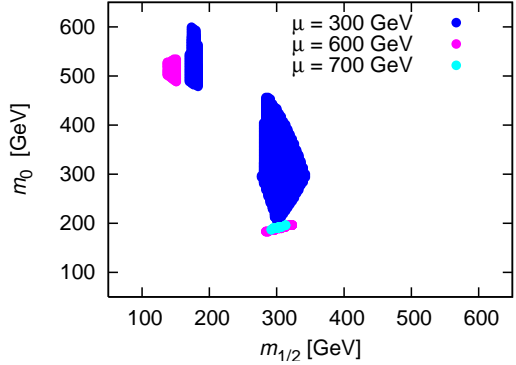}} \ \
\caption{The regions are allowed
by the recent results for the dark matter abundance,
$b\to s\gamma$, and $B_s \to \mu^+\mu^-$
with
$(\mu / {\rm \,GeV},A_0 / {\rm \,GeV}) = (300, -700)$ (blue (black)),
$(600, -1000)$ (magenta (gray)), $(700, -1100)$ (light blue (light gray)).
 The regions are searched for $m_A = 80 \text{-} 140\,\GeV$.
}
\label{fig:mu}
\end{center}
\end{figure}
%
\begin{figure}[t]
\begin{center}
\scalebox{0.9}{\includegraphics*{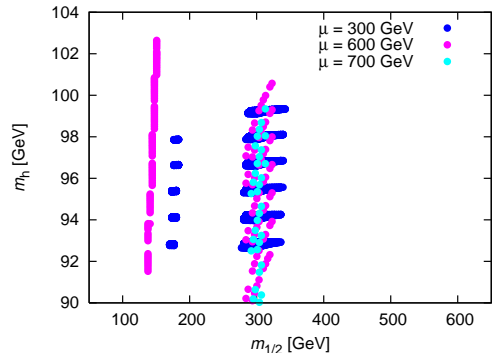}}
\caption{
 The mass $m_h$ of the lightest CP-even scalar Higgs
for the allowed regions in Fig.\,\ref{fig:mu}.
}
\label{fig:mh}
\end{center}
\end{figure}

In Fig.\,\ref{fig:mu},
we show the regions allowed by $\Omega_{\text{DM}} h^2$,
${\rm Br}(b \to s \gamma )$, and ${\rm Br} ( B_s \to \mu^+ \mu^-)$
in $(m_{1/2}, m_0)$ plane
for $m_A = 80 \text{-} 140\,\GeV$.
The allowed regions are obtained for
$(\mu / {\rm \,GeV},A_0 / {\rm \,GeV}) = (300, -700)$ (blue (black)),
$(600, -1000)$ (magenta (gray)), $(700, -1100)$ (light blue (light gray)).
For small $\mu$, the large allowed region is derived and
the portion around $m_{1/2}\sim 300$\,GeV is mild bino-Higgsino mixing region.
If the mixing is large, the LSP pair can intensely annihilate
and the relic abundance is too small.
On the other hand, the allowed region is very restricted for large $\mu$.
 Figure~\ref{fig:mh} shows values of $m_h$
for the allowed regions presented in Fig.\,\ref{fig:mu}.
 It is clear that a wide range of $m_h$ bellow $114.4\,\GeV$ is allowed.

The spin-independent neutralino-nucleon elastic-scattering cross section
for direct detection experiments is large
for small masses of the neutral Higgs bosons~\cite{LHSBottino1,DirectLHS}.
In the LHS, because all Higgs boson masses are as light as the electroweak scale,
the prediction for the cross section is generically large.
Thus, we should compare the prediction
with the bound from direct detection experiments.
%
\begin{figure}[t]
\begin{center}
\scalebox{.95}{\includegraphics*{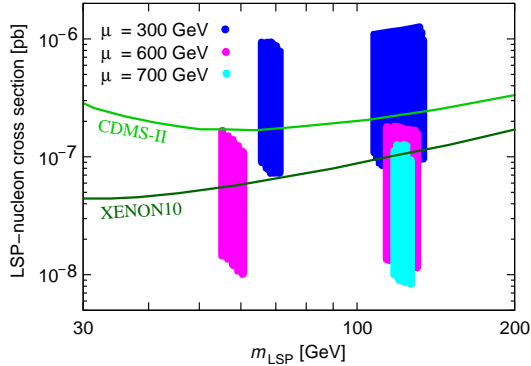}}
\caption{The predictions for direct detection experiments
of the dark matter are shown
for the same parameter sets as Fig.\,\ref{fig:mu}.
 The regions are obtained for $y = 0 \text{-} 0.2$.
See the text for the detail.
}
\label{fig:directdetection}
\end{center}
\end{figure}
%
 In Fig.\,\ref{fig:directdetection}, 
the predictions for direct detection experiments are shown
for the same parameter sets as Fig.\,\ref{fig:mu}.
The horizontal axis is the LSP mass and
the vertical one is the spin-independent cross section
for direct detection experiments.
The cross section is calculated for $y=0 \text{-} 0.2$,
where $y$ parametrizes the strangeness component in nucleons,
to take into account that the cross section becomes smaller for smaller $y$.
The value of $y$ is calculated as $0.2$ in Ref.\,\cite{Gasser:1990ce},
while $y=0$ is also allowed within errors.
The dark green (dark gray) and green (gray) lines indicate the bounds
from the XENON10 and 
the CDMS-II 
experiments~\cite{DMDirect}, respectively.
The small value of $\mu $ is disfavored
even if it gives large allowed region in Fig.\,\ref{fig:mu}.
For $\mu = 600, 700\,\GeV$,
the cross section is consistent with experiments.
Since the allowed region with the WMAP observed dark matter abundance
is very restricted for $ \mu = 700\,\GeV$ in Fig.\,\ref{fig:mu},
moderate value of $\mu$ is preferable in the LHS\@.
The regions allowed by direct detection experiments
are not far away from recent bounds of the experiments,
and hence the LHS predicts that the dark matter
may be observed by experiments in near future.

\begin{center}
 {\bf V. \ SUMMARY}
\end{center}
 We have investigated phenomenology of the neutralino in the LHS
for $90\,\GeV < m_h < 114.4\,\GeV$.
We confirmed that the abundance of the neutralino is consistent
with the observed dark matter abundance.
The constraint from ${\rm Br} (b \to s \gamma)$
severely restrict a part of the region allowed by the abundance.
However, we can choose $A_0$ to cancel the dangerous contribution
from $H^\pm$ to $b \to s \gamma$.
 The recent constraint on $ {\rm Br} (B_s \to \mu^+ \mu^-)$
is avoided for small $\tan \beta $
and the large part of the predicted value is within the sensitivity
in future experiments.
It was also shown that the allowed region is very restricted for large $\mu$.
In the LHS, the cross section for direct detection experiments
of dark matter is large.
In particular, small $ \mu $ is disfavored by the present bound.
Hence, moderate value of $\mu$, 600\,GeV or so, is favored in the LHS\@.
 The allowed region may be judged by $B_s \to \mu^+ \mu^-$
and direct detection experiments in near future.\\

{\bf Acknowledgments:}
We thank J.~Hisano and K.~Tobe for valuable comments.
The work of MS is supported in part by the Grant-in-Aid for Young Scientists.

\end{document}